\documentstyle[aps,epsf,prl,floats,twocolumn]{revtex}
\begin{document}
\draft
\tighten
\title{Double beta decay of $^{48}$Ca}
\author{A. Balysh$^1$, A. De Silva$^2$, V. I. Lebedev$^1$, K. Lou$^3$, 
M. K. Moe$^2$,\\ M. A. Nelson$^{2}$\cite{mattnow}, A. Piepke$^3$, 
A. Pronskiy$^1$, 
M. A. Vient$^{2}$\cite{mikenow}, and P. Vogel$^3$}
\address{$^1$ Kurchatov Institute, Kurchatov Square, 123182 Moscow, Russia \\
$^2$ Department of Physics and Astronomy, University of California Irvine, 
Irvine, California 92697, USA \\
${}^3$ Physics Department, California Institute of Technology,
Pasadena, California 91125, USA}
\date{\today}
\maketitle

\begin{abstract}
$^{48}$Ca, the lightest double beta decay candidate,
is the only one simple enough to be treated exactly in the nuclear shell 
model. Thus,
the $\beta\beta(2\nu)$ half-life measurement, reported here, provides 
a unique test
of the nuclear physics involved in the $\beta\beta$ matrix element calculation.
Enriched $^{48}$Ca sources of two different thicknesses have been exposed 
in a time
projection chamber, and yield T$_{1/2}^{2\nu} = (4.3^{+2.4}_{-1.1}
[{\rm stat.}] \pm 1.4 [{\rm syst.}]) \times 10^{19}$ years,
compatible with the shell model calculations.
\end{abstract}


\maketitle

Among all candidates for double beta decay, 
$^{48}$Ca$\rightarrow ^{48}$Ti is unique,
since it is the only one which can be treated ``exactly" in the nuclear shell
model by  solving the problem of eight nucleons
distributed within the $fp$ shell without truncation.
Consequently, this decay has been 
a favored testing ground of nuclear theories \cite{BV}. 
However, until now
only a lower limit of the $2\nu$ decay half-life, 
T$_{1/2}^{2\nu} \geq 3.6 \times 10^{19}$ years,
has been determined experimentally \cite{wu}.

In a recent paper \cite{radha} it was shown that 
the nuclear
shell model, constrained by the requirement that it describes well the 
spectroscopy of the $A=48$ nuclei, restricts the corresponding half-life
also from above, 
T$_{1/2}^{2\nu} \leq 10^{20}$ y.  
Therefore the experimental
observation of the decay acquired added significance; if it turns out
that the shell model cannot predict this theoretically tractable rate,
we have to wonder about our ability to describe the nuclear matrix
elements in the more complex nuclei, including perhaps even
the matrix elements responsible for 
the neutrinoless decay. 
Here we will present the result from a new experiment using a time 
projection chamber (TPC).

The search for double beta decay of $^{48}$Ca was among the first to be 
attempted in live-time experiments beginning in the early fifties 
(for an extensive chronology see \cite{HS}).  
With the largest energy release of all $\beta\beta$ candidates 
$^{48}$Ca (Q$_{\beta\beta}$ = 
4.3~MeV) has a $\beta\beta(2\nu)$ sum-energy spectrum that extends to higher 
energies than most radioactive background.  Yet calcium has a
tendency to harbor chemically similar radio-impurities such as $^{90}$Sr and 
$^{226}$Ra, which do intrude on a major fraction of the $^{48}$Ca spectral 
range.  

When the two $\beta$ particles are tracked in a TPC 
they are seen to both carry negative charge, originate from a common point, and 
have separately measured energies.  Although this distinctive 
visualization eliminates the bulk of unrelated activity,
there remain several well-known mechanisms for production of negative 
electron pairs that constitute background for $\beta\beta$ decay.
The most serious of these are M\"{o}ller scattering of single $\beta$ 
particles, and $\beta$-$\gamma$ cascades in which a $\gamma$-ray internally 
converts or Compton scatters.
These processes are fed principally by the primordial decay chains, but also
by cosmogenic and man-made radionuclides.  Decay-chain induced background 
events can often be tagged by $\alpha$ or $\beta$ particles from neighboring 
links if the source is thin enough to allow the tagging particles escape 
into the TPC gas.

The miniscule 0.187\% natural abundance of $^{48}$Ca makes 
enrichment both necessary and expensive.  Potential exposure to 
loss of costly isotope is a deterrent to chemical purification 
or conversion to the lightest stable compound.
The material used was supplied by the Kurchatov Institute as 
finely powdered CaCO$_{3}$ enriched to 73\% 
in $^{48}$Ca, and relatively free of U and Th ($<0.8$ ppb by mass spectroscopic 
analysis).  The speculative assumption of secular equilibrium and the 
absence of serious amounts of other high Q-value impurities, led to 
a projected experimental $\beta\beta$ sensitivity in excess of $10^{20}$ years for
this material.
Accordingly, a relatively thick $\beta\beta$ source was made to maximize the number 
of exposed $^{48}$Ca nuclei.  The CaCO$_{3}$ powder was injected into a 
large glass box in bursts of compressed gas, allowed to settle 
uniformly onto a 4 $\mu$m Mylar substrate, then fixed with a mist of 
Formvar.  Two such deposits, face-to-face, formed the first $\beta\beta$ 
source, with a total of 42.2 g of CaCO$_{3}$ (18.5 
mg/cm$^{2}$ total thickness with substrate and binder.)

The shielded UC Irvine TPC \cite{NIM88,Erice94} containing 
the $\beta\beta$ source as the central electrode in a magnetic field was 
located in a tunnel at the 
Hoover Dam under a minimum of 72 m of rock.  Data were recorded on 
magnetic tape, and subsequently passed through stripping software to 
select clean 1e$^{-}$ and 2e$^{-}$ events.  The 2e$^{-}$ events were 
individually scanned by a physicist.  All unambiguous negative pairs 
emitted from opposite sides of the source with a common point of 
origin were fitted with helices, and the parameters written to a $\beta\beta$ 
candidate file or a $^{214}$Bi file, depending on whether a $^{214}$Po 
$\alpha$ particle appeared at the vertex within the following millisecond.
The far more numerous 1e$^{-}$ events were fitted automatically 
by software and also written to a parameter file.

\begin{figure}[htb]
\vspace{-0.5cm}
\hspace{0.5cm}
\mbox{\epsfxsize=5.5 cm\epsffile{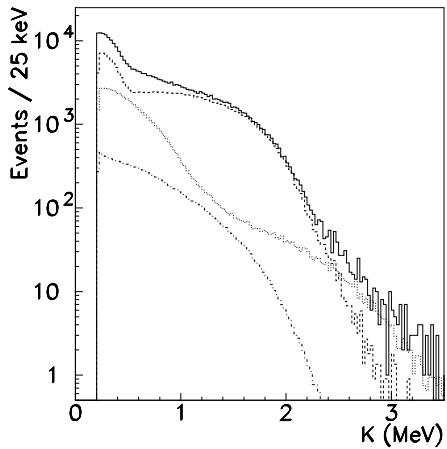}}
\vspace{-0.5cm}
\caption{The lone electron spectrum for the thick source, and the most 
important of the
fitted components.  From top to bottom at low energy are
the total spectrum, $^{90}$Sr, $^{226}$Ra, and $^{228}$Ra.
Daughters are assumed in equilibrium, with the exception of $^{210}$Pb.  
Its daughter $^{210}$Bi was fitted separately, then included in 
the $^{226}$Ra curve.}
\label{fig:Lone}
\end{figure}

The lone electron (1e$^{-}$) spectrum plotted against kinetic energy~$(K)$ 
in Fig. \ref{fig:Lone} represents 
the total 
beta activity of source contaminants, and can be broken down 
into the contributions from individual radionuclides by a least squares
fit.  $^{90}$Sr (2250 $\mu$Bq/g), $^{226}$Ra (530 $\mu$Bq/g) 
and their daughters account 
for the bulk of the spectrum.  Contributions from $^{137}$Cs 
(940 $\mu$Bq/g), 
and daughters of $^{228}$Ra (90 $\mu$Bq/g), are also present.  
The two Ra activities being much larger than the mass spectroscopic 
limits on U and Th, indicate severe breaking of equilibrium in the 
respective series.  The 
daughters of greatest concern (Q$_{\beta}>2$ MeV) are 
$^{90}$Y, 
$^{214}$Bi, 
$^{228}$Ac, 
$^{212}$Bi, 
and $^{208}$Tl.  As a check of the fitting procedure, the energy spectrum 
of electrons tagged by $^{214}$Po $\alpha$-particles was noted to closely 
match the fitted $^{214}$Bi component when 
adjusted for the $\alpha$ escape probability (P$_{\alpha}=0.24\pm 0.01$ 
from an independent measurement.)

The activities determined from the lone electron spectrum were used as 
input to a Monte Carlo calculation of the 2e$^{-}$ background.  $^{90}$Sr and 
its daughter $^{90}$Y are essentially pure $\beta$ emitters, and contribute 
only through M\"{o}ller scattering.  The other high Q$_{\beta}$ nuclei have 
complex decay schemes with multiple gamma rays \cite{Lederer}, all of 
which were included in the Monte Carlo with their corresponding
conversion coefficients.  Most of these simulated 2e$^{-}$ backgrounds 
were directly testable against TPC measurements, with good agreement:
The measured $^{214}$Bi component was simply the 2e$^{-}$ data subset tagged 
by the $^{214}$Po $\alpha$, and corrected for the $\alpha$ escape probability.
The $^{90}$Y\, 2e$^{-}$ measurement was provided by a drop of 
$^{90}$Sr solution applied to a natural 
isotopic replica of the $^{48}$Ca source, and  placed in the TPC.
The 
$^{212}$Bi and 
$^{208}$Tl 2e$^{-}$ measurements 
were scaled from those produced by an injection of $^{220}$Rn, by 
comparing observed rates of the rapid $^{212}$Bi-$^{212}$Po, 
$\beta$-$\alpha$ sequence.

Since the Monte Carlo 2e$^{-}$ rates were derived from intrinsic 
activity levels in $\mu$Bq/g, their agreement with direct measurements also 
confirms the Monte-Carlo predicted 2e$^{-}$ efficiency of the TPC.
A Monte-Carlo generated background spectrum was essential only for $^{228}$Ac 
where we have no 
TPC measurement, but in our 2e$^{-}$ background model we elected to 
use the smoother, better-statistics Monte Carlo spectra for the 
other contributions as well.

An alternative determination of 2e$^{-}$ background was carried out by a 
separate subgroup of the collaboration, and included independent Monte 
Carlo calculations and a greater reliance on the above-mentioned 
TPC measurements as opposed to lone-electron fits.  We refer 
to this direct measurement method as analysis ``A'', and the 
lone-electron based method as analysis ``B''.  To eliminate events 
with the poorest energy resolution, analysis B 
included a cut on 
electrons making the smallest angles with the magnetic field,
($\mid\! \cos(\theta)\!\mid\: < 0.9$).

\begin{figure}[htb]
\vspace{-1.0cm}
\hspace{0.5cm}
\mbox{\epsfxsize=5.5 cm\epsffile{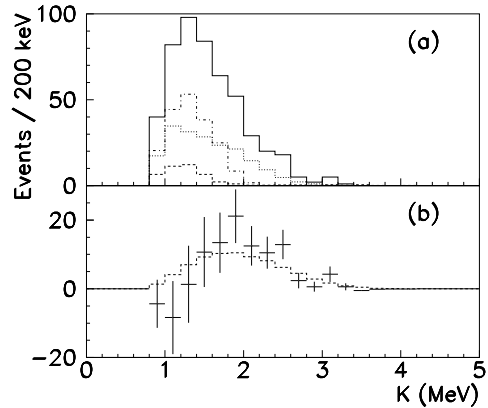}}
\caption{(a) The measured 2e$^{-}$ sum-energy spectrum 
and background spectra
for the thick source with a 400 keV singles 
threshold.  From top to bottom at low energy are the total 
measured spectrum, M\"{o}ller events, $^{214}$Bi, and daughters of 
$^{228}$Ra.
(b) Residual $\beta\beta$ candidates and Monte Carlo $\beta\beta$ 
spectrum at the corresponding 
T$_{1/2}^{2\nu}$ = $4.1\times 10^{19}$ y obtained at this
threshold (dashed line).}
\label{fig:sum}
\end{figure}

The 2e$^{-}$ sum spectrum from the $^{48}$Ca source, after event-by-event 
removal of the $\alpha$-tagged $^{214}$Bi, is shown in Fig. \ref{fig:sum}a 
with a singles threshold of 400 keV.  
The 
various remaining background spectra as determined by analysis~A are 
superimposed.  The residual 
spectrum following background subtraction, appears in Fig. 
\ref{fig:sum}b.

The above procedure was repeated for a series of singles 
thresholds and one relatively high sum threshold, with consistent results.  The 
corresponding half-lives were calculated in each case, as shown for 
two of the threshold combinations in 
Table \ref{tab:TABLE}.  The 2 MeV sum threshold was accompanied by an 
additional singles spectrum cut on the
strong $^{214}$Bi conversion line at 1.3 MeV.
Quoted errors in the table are statistical.
The calculated half-life 
is independent of threshold and analysis A or B within errors.

Since the half-life precision was degraded by the unexpectedly large Ra
and Sr 
contamination, the 42.2 g source was replaced after 2440 hours
exposure, with one containing only 10.3 g of enriched CaCO$_{3}$, 
and total thickness 5.4 mg/cm$^{2}$.  
The thinner source was exposed 4001 hours, and these results are also 
included in Table \ref{tab:TABLE}.  
The $^{214}$Bi rejection improved greatly as a result of an increase in 
P$_{\alpha}$ from 0.24 to 0.69, and 
the $^{90}$Y component was weakened by 
the large reduction in target mass 
for M\"{o}ller scattering.

In view of these significant improvements, the relatively small increase 
in signal-to-background ratio observed for the thin source was 
disappointing.  We now understand that multiple Coulomb
scattering caused a nonlinear 
dependence of total M\"{o}ller cross section on source thickness.
In addition, the thick source degraded a larger percentage of electrons 
to sub-threshold 
energies, and thereby suppressed background more effectively than the 
generally more energetic $\beta\beta$ events.
Nevertheless, it is encouraging that despite different dependence on 
source mass for the $^{48}$Ca signal and the background rates 
(particularly $^{214}$Bi,) the half-lives derived from thick and thin 
sources agree within statistics.

\begin{figure}[htb]
\vspace{-0.5cm}
\hspace{0.5cm}
\mbox{\epsfxsize=5.5 cm\epsffile{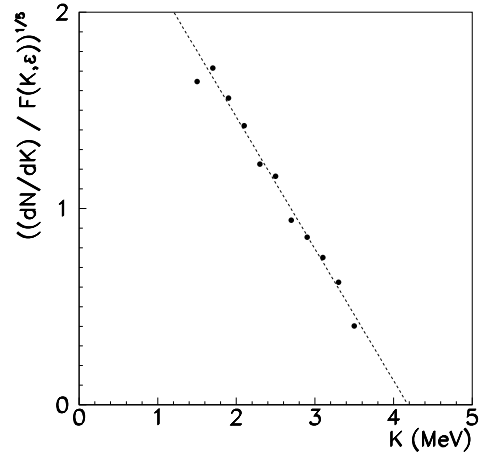}}
\caption{High-energy portion of the Kurie plot
for residual $\beta\beta$ 
candidates combined from thick and thin sources
at a singles threshold of 400 keV.  
($F(K,\epsilon)=(K-2\epsilon)(f_0(K)+f_{\epsilon}(K))$. See text.)}

\label{fig:kurie}
\end{figure}

The thick and thin source residual spectra were each 
corrected for energy loss and efficiency distortion, then combined 
for the Kurie plot shown in Fig. \ref{fig:kurie}.  
Including the singles threshold ($\epsilon$) in an integration of the 
Primakoff-Rosen spectral approximation \cite{Prim} produced
a Kurie formula that retains the linear energy dependence of a transformed 
$\beta\beta(2\nu)$ spectrum in the presence of a singles threshold 
cut.
The resulting formula is then
\[\left[ \frac{dN}{dK}/((K-2\epsilon)(f_0(K)+f_{\epsilon}(K))) \right]^{1/5} 
\propto \; (Q_{\beta\beta} - K) \]
where
\[f_0(K) ~=~ K^4/30 + K^3m/3 +4K^2m^2/3 + 2Km^3 + m^4 \]
and
\[f_ {\epsilon}(K) ~=~ {\epsilon}(K - {\epsilon})(K^2/15 + 2Km/3 +2m^2/3 
+ {\epsilon}(K-{\epsilon})/5) ~.\]
with electron mass ($m$).
The Kurie plot
energy intercept at $4.2\pm 0.1$ MeV is consistent with the 
$^{48}$Ca Q$_{\beta\beta}$ 
value of $4.271\pm 0.004$ MeV \cite{Wapstra}.
The small error bars resulting from the Kurie 
transformation, have been omitted in the figure.
By comparison, the $\alpha$-tagged $^{214}$Bi 2e$^{-}$ events 
produce a distinctly bowed 
Kurie plot, and the plot for 
measured $^{208}$Tl (Q$_{\beta}=5.0$ MeV) events 
is grossly nonlinear.

Since the thick and thin sources were exposed separately,
corresponding pairs of columns in the table can be combined as independent
measurements.  For example, combining thick and thin $\beta\beta$ events for
0.400/0.800 MeV singles/sum thresholds, analysis A, yields
T$_{1/2}^{2\nu} = (4.3^{+2.4}_{-1.1})\times 10^{19}$ y.
Either of the other two thick-thin pairs would combine to give an equally valid 
result.  However, since the other two results would not be statistically 
independent from the first, we do not attempt a grand average.  Rather 
we choose the 
above number, since it includes a broader range of the spectrum than 
the higher sum threshold, and we include the A-B analysis difference 
in the systematic error.  The remainder of the systematic error is
largely in the detector efficiency.  Thus we quote a final result of
\mbox{T$_{1/2}^{2\nu} = (4.3^{+2.4}_{-1.1}
[{\rm stat.}] \pm 1.4 [{\rm syst.}]) \times 10^{19}$ y.}

\begin{table*}[htb]
\caption{\protect Breakdown of counts from the two sources for two
energy thresholds.}
\label{tab:TABLE}
\renewcommand{\arraystretch}{1.2}
\begin{tabular}{lcccccc} \\ 
\multicolumn{1}{l}{} & \multicolumn{3}{c}
{\underline{~~~~~~~~Thick Source (0.0775 mol$\cdot y$)~~~~~~~~}}
&\multicolumn{3}{c}{\underline{~~~~~~~~Thin Source (0.0310 
mol$\cdot y$)~~~~~~~~}} \\
 &&&&&& \\
Singles threshold (MeV) &\multicolumn{2}{c}{0.400} &0.200
&\multicolumn{2}{c}{0.400} &0.200 \\
Sum threshold (MeV) &\multicolumn{2}{c}{0.800} &2.000 
&\multicolumn{2}{c}{0.800} &2.000 \\
$^{214}$Bi 1.3 MeV line cut &\multicolumn{2}{c}{No} &Yes 
&\multicolumn{2}{c}{No} &Yes \\ \hline
 &\underline{Analysis A} &\underline{Analysis B} &
&\underline{Analysis A} &\underline{Analysis B} & \\

 &&&&&& \\
Small polar angles cut &No &Yes &No &No &Yes &No \\
Counts &&&&&& \\
~~~~With $\alpha$ tag &55 &50 &$6.8\pm 1.4$ \tablenotemark[1]
 &79 &72 &$8.0\pm 1.6\: $\tablenotemark[1] \\
~~~~Without $\alpha$ tag &500 &472 &72 &142 &134 &21 \\
Backgrounds \tablenotemark[2]&&&& \\
~~~~Untagged $^{214}$Bi &$189.1\pm 25.5$ &$186.3\pm35.7$ 
&$23.4\pm 4.9$ &$40.2\pm 4.5$ 
&$36.4\pm8.9$ &$4.1\pm 0.8$\\
~~~~M\"{o}ller events \tablenotemark[3]
 &$191.5\pm 4.6$ &$163.4\pm5.6$&$3.2\pm 0.6$ 
&$53.7\pm 1.7$ &$46.1\pm 2.1$ &$1.5\pm 0.3$ \\
~~~~$^{228}$Ac, $^{212}$Bi, $^{208}$Tl \tablenotemark[4]
&$42.8\pm 7.5$ 
&$29.4\pm5.6$ &$4.0\pm 0.8$ &$19.8\pm 3.5$  &$13.3\pm2.1$ &$1.4\pm 0.3$ \\ 
Total background &$423.3\pm 27.0$ &$379.1\pm36.6$&$30.7\pm 5.0$ 
&$113.7 \pm 6.0$ 
&$95.8\pm9.4$&$7.0\pm 0.9$ \\ \hline
$\beta\beta$ events &$76.7\pm 35.0$ &$92.9\pm42.5$ 
&$41.3\pm 9.8$ &$28.3\pm 13.3$ 
&$38.2\pm14.9$&$14.0\pm 4.7$ \\
Efficiency \tablenotemark[5]&0.0973 &0.0902 &0.0414 
&0.1046 &0.0972 &0.0469 \\
T$^{2\nu}_{1/2}$ $(10^{19}$ y) &$4.1^{+3.5}_{-1.3}$ 
&$3.2^{+2.7}_{-1.0}$&$3.3^{+1.0}_{-0.6}$ &$4.8^{+4.2}_{-1.5}$  
&$3.3^{+2.1}_{-0.9}$&$4.3^{+2.2}_{-1.1}$  \\
Signal/Background &0.18 &0.25 &1.35 &0.25 &0.40 &2.04 \\ 
Kurie plot intercept (MeV) &$4.1\pm 0.1$ &&&$4.4\pm 0.1$ 
&&
\end{tabular}
\tablenotetext[1]{Scaled from counts at lower
threshold by the ratio observed in a larger $^{214}$Bi data set.}
\tablenotetext[2]{See text.}
\tablenotetext[3]{Exclusive of $^{214}$Bi which is included in the row above.}
\tablenotetext[4]{Exclusive of M\"{o}ller,which is included in the row above.} 
\tablenotetext[5]{From Monte Carlo simulation.}
\end{table*}

This observation is consistent with the spectroscopy-constrained shell model
of reference \cite{radha}.
For most $\beta\beta$ nuclei the agreement between 
the experimental half-life and the calculated one has been obtained 
only with the empirical quenching of the Gamow-Teller 
(GT) strength, i.e., the multiplication of all matrix elements of the 
operator $\sigma\tau_{\pm}$
by a universal quenching factor 0.77.
Taking into account
the range of lifetimes compatible with the limits of Ref. \cite{radha}, and
the experimental uncertainties described above, we cannot, alas, make 
a definitive statement about the GT quenching for the $2\nu$ double
beta decay (but for the optimal shell model the quenching is clearly required). 
On the other hand, some calculations \cite{muto} that are 
unconstrained by the spectroscopy test, suggest a
much longer half-life for $^{48}$Ca in disagreement with the present 
experiment.

Although the $^{48}$Ca enriched sample contained traces of radium and strontium 
activity,
backgrounds were well defined by associated 
alpha particles and the lone electron spectrum.  The residual data give 
half-life values that are consistent between two $\beta\beta$ sources of 
different thickness, and among various energy thresholds.  The 
corresponding Kurie plots all intercept the energy axis near the 
$^{48}$Ca Q$_{\beta\beta}$ value, unlike plots made from measured 
samples of the various backgrounds.  We believe these results 
constitute strong evidence for $^{48}$Ca double beta decay at a half-life 
supporting the relatively rigid shell model calculations for this 
lightest double beta decay nucleus.

We gratefully acknowledge the hospitality of 
\linebreak
Blaine~Hamann
and William~Sharp at the Hoover Dam.  We thank Felix~Boehm for his advice 
and support, and one of us (A.P.) wishes to acknowledge the support of
the Alexander von~Humboldt Foundation.
This work was funded by the US Department of Energy under contracts 
DE-FG03-01ER40679 and DE-FG03-88ER40397.



\begin{references}
%
\bibitem[*]{mattnow} Present address: Dynamics Technology, Inc.,
21311 Hawthorne Blvd., Suite 300,
Torrance, CA 90503-5610.
\bibitem[\dag]{mikenow} Present address: 3901 E. Desert Flower Lane,
Phoenix, AZ  85044.
%
\bibitem{BV} For a representative list 
see F. Boehm and P. Vogel, {\it Physics of Massive Neutrinos},
Cambridge: Cambridge Univ. Press. 2nd ed. (1992)
\bibitem{wu} R. K. Bardin, P. J. Gollon, J. D. Ullman and C. S. Wu,
Nucl. Phys. {\bf A158}, 337 (1970).
\bibitem{radha} A. Poves, R. P. Bahukutumbi, K. H. Langanke and P. Vogel,
Phys. Lett. {\bf B361}, 1 (1995).
\bibitem{HS}  W. C. Haxton and G. J. Stephenson, Jr,
Progr. Part. Nucl. Phys. {\bf 12}, 409 (1984).

\bibitem{NIM88} S. R. Elliott, A. A. Hahn and M. K. Moe, Nucl. 
Instrum. Meth. in Phys. Res., {\bf A273}, 226 (1988).
\bibitem{Erice94} M. K. Moe, M. A. Nelson and M. A. Vient, Prog. Part. 
Nucl Phys. {\bf 32}, 247 (1994).
\bibitem{Lederer} {\it Table of Isotopes --- 7th Edition}, 
edited by C. M. Lederer 
and V. S. Shirley (John Wiley \& Sons, New York, 1978).
\bibitem{Prim} H. Primakoff and S. P. Rosen, Rep. Prog. Phys. {\bf 22}, 
121 (1959).
\bibitem{Wapstra} A. H. Wapstra and G. Audi, Nucl. Phys. {\bf A432}, 1 (1985).
\bibitem{muto} K. Muto, E. Bender and H. V. Klapdor-Kleingrothaus,
Z. Phys. {\bf A339}, 435 (1991); 
M.K.Cheoun, A. Bobyk, A. Faessler, F. Simkovic, and G. Teneva,
Nucl. Phys. {\bf A561}, 74 (1993).

\end{references}
\end{document}